\documentclass[fleqn,12pt,twoside]{article}
\usepackage{espcrc1}

\usepackage{graphicx}
\usepackage[figuresright]{rotating}

\newcommand{\AmS}{{\protect\the\textfont2
  A\kern-.1667em\lower.5ex\hbox{M}\kern-.125emS}}
\def\be{\begin{eqnarray} &&}
\def\nonu{\nonumber \\ &&}
\def\ee{\end{eqnarray}}

\title{Timelike and spacelike hadron form factors, 
Fock state components and light-front dynamics}

\author{ E. Pace\address{Dipartimento di Fisica, Universit\`a di Roma "Tor Vergata" 
and Istituto Nazionale di Fisica Nucleare, Sezione Tor Vergata, Via della Ricerca
Scientifica 1, I-00133  Roma, Italy},
G. Salm\`e\address{Ist. Nazionale di Fisica Nucleare, Sezione Roma I, P.le A. Moro 2,
 I-00185 Roma, Italy}
       ,
        T. Frederico\address{Dep. de F\'\i sica, Instituto Tecnol\'ogico da Aeron\'autica,
Centro T\'ecnico Aeroespacial, 12.228-900 S\~ao Jos\'e dos
Campos, S\~ao Paulo, Brazil},
       S. Pisano\address{Dip. di Fisica, Universit\`a di Roma "La Sapienza", P.le A. Moro 2,
 I-00185 Roma, Italy}
        and
	J. P. B. C. de Melo\address[MCSD]{Centro de Ci\^encias Exatas e Tecnol\'ogicas,
 Universidade Cruzeiro do Sul, 08060-070,  and
Inst. de F\'\i sica Te\'orica, Universidade Estadual Paulista
 01405-900, S\~ao Paulo, Brazil}%
 }
       
\begin{document}

\maketitle
\begin{abstract}
A unified description of spacelike and timelike hadron form factors within a light-front model
was successfully applied to the pion. The model is extended to the nucleon
 to study the role of $q \bar q$ pair production and of nonvalence components
 in the nucleon form factors. Preliminary results in the spacelike range
$0 \le Q^2 \le 10 ~ (GeV/c)^2$ are presented.  
\end{abstract}

\vspace{0.3cm}

The controversial experimental results for the nucleon form factors (FF) in the
spacelike (SL) region \cite{JLAB} and the striking differences between 
data \cite{Fenice} and
 theoretical expectations from perturbative QCD \cite{Karliner} 
in the timelike (TL) region motivate
a unified investigation of SL and TL
nucleon FF for a better knowledge of the nucleon internal structure.

Within the light-front dynamics \cite{brodsky}, we developed an approach 
for a global description of the pion FF in both the SL and
 the TL regions, which is able to reproduce successfully the experimental data 
from $q^2 = - 10~(GeV/c)^2$ up to $q^2 = 10~(GeV/c)^2$ \cite{DFPS}.
We extend now the procedure already applied in the pion case
to the nucleon, considered as a system of constituent quarks 
of mass  $m = .200~GeV$.
As a first step,
we study the nucleon FF in the SL region. In particular we aim to investigate the
relevance of the contribution due to the $q\bar{q}$ pair creation
 by the incoming virtual photon.

For a unified description of the nucleon FF in both the SL and the TL regions 
a reference frame with $q^+ \ne 0$ is needed. Therefore, following Ref. \cite{LPS}, 
 we calculate 
the nucleon form factor in the SL region in the reference frame
 where ${\bf q}_{\perp}=0$ and $q^+ = [{Q^2}]^{1/2}$. 

 To introduce a proper Dirac structure 
for the nucleon Bethe-Salpeter amplitude (BSA), we describe the $qqq$-nucleon interaction
through an  effective Lagrangian, which  represents an isospin zero,
 spin zero coupling for the (1,2) quark pair, as in Ref. \cite{nua} but with $\alpha = 1$. 

Then, the nucleon Bethe-Salpeter amplitude  is approximated as follows
\be
\Phi^{\sigma}_N(k_1,k_2,k_3,p) 
=\left [~S(k_1)~\imath \tau_y ~ \gamma^5 ~ S_C(k_2)C ~ S(k_3)~+
~S(k_3)~\imath \tau_y ~ \gamma^5 ~S_C(k_1)C ~ S(k_2)~+
\right. \nonu 
\left. ~S(k_3)~\imath \tau_y ~ \gamma^5 ~S_C(k_2)C~S(k_1)~
 \right ] 
 ~\Lambda(k_1,k_2,k_3)~\chi_{\tau_N}~U_N(p,\sigma)
 \label{ampli1}
\ee
where $\Lambda(k_1,k_2,k_3)$ describes  the symmetric momentum dependence of the
vertex function upon the quark momentum variables, $k_i$. 
In the Bethe-Salpeter amplitude (\ref{ampli1}), 
 $U_N(p,\sigma)$ is the nucleon  Dirac spinor
and $\chi_{\tau_N}$ the nucleon isospin state, while  the quantities
$ S(k)$ and $S_C(k)$
are the propagators for the quark and the charge conjugated quark, respectively.

The matrix elements of the {\em macroscopic} em nucleon current,
 \begin{equation}
 \langle p',\sigma'|j^\mu|p,\sigma \rangle = 
 \bar U_N(p',\sigma') \left [-F^N_2(Q^2)  { {p'}^\mu +{p}^\mu \over 2M_N}
+\left (F^N_1(Q^2)+F^N_2(Q^2)\right )\gamma^\mu\right] U_N(p,\sigma) \quad ,
\label{spc2}
\end{equation}
where $F^N_{1(2)}(Q^2)$ is the Dirac (Pauli) nucleon FF, 
in impulse approximation
can be approximated {\em microscopically} by the Mandelstam formula \cite{mandel} as follows
\be
\langle  \sigma',p'|j^\mu~|p,\sigma \rangle =
N_c ~ \int {d^4k_1 \over (2\pi)^4}\int {d^4k_2 \over (2\pi)^4} \times \nonu
 Tr_{\tau_{(1,2)}} ~Tr_{\tau_{(3,N)}}~Tr_{\Gamma_{(1,2)}}~
\left \{~\bar \Phi^{\sigma'}_N(k_1,k_2,k'_3,p')~S^{-1}(k_1)~S^{-1}(k_2)~{\cal I}^\mu _3~
 ~\Phi^\sigma_N(k_1,k_2,k_3,p)\right \}
\label{spc1}
\ee
where $N_c$ is the number of colors, the traces are performed over Dirac ($\Gamma$) and
isospin ($\tau$) indexes,
the subscripts in the traces indicate the particles involved, and
${\cal I}^\mu _3$ is the quark-photon vertex for the third quark.
 The  quark-photon vertex has isoscalar 
and isovector contributions, namely ~~
 ${\cal I}^\mu_3 =
  ~({\cal I}^\mu_u~+~{\cal I}^\mu_d )/ 2~+~\tau_z~({\cal I}^\mu_u~-~{\cal I}^\mu_d )/ 2 ~
  =~{\cal I}^\mu_{IS} +\tau_z ~{\cal I}^\mu_{IV}$  ~.
 
 \begin{figure}
\includegraphics[width=14.cm]{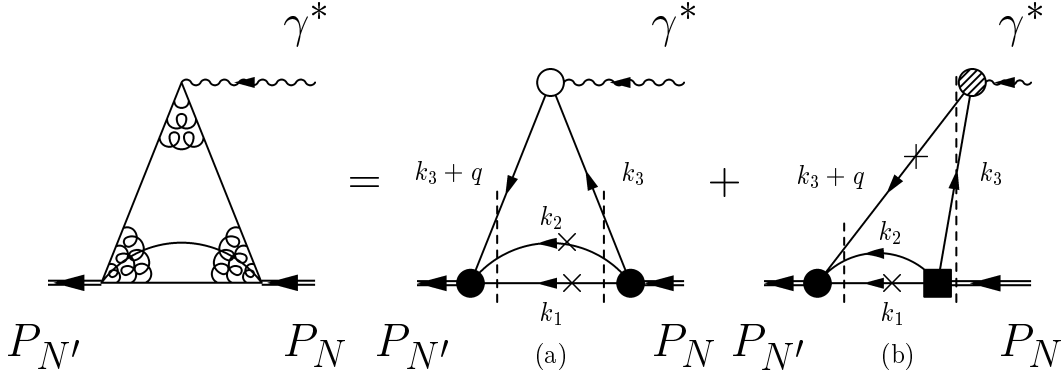}
\caption{Diagram (a) : valence, triangle contribution  ($0 < k_{i}^+ < P^+_{N}$,
$0 < k_{3}^+ + q^+ < P^+_{N^\prime}$).
 Diagram (b) : nonvalence contribution  ($0 > k_3^+ > - q^+$). The simbol $\times$ 
 on a quark line indicates a quark on the mass shell, i.e.
$ ~ k^-_{on} = (m^2 + k^2_{\perp})/k^+$.}
\end{figure}
We assume a suitable fall-off of the momentum components of the 
  BSA $\Lambda(k_1,k_2,k_3)$ and $\Lambda(k_1,k_2,k'_3)$, 
   to make finite the four dimensional integrations  in Eq. (\ref{spc1}).
  Furthermore, we assume that the singularities of 
   $\Lambda(k_1,k_2,k_3)$ and $\Lambda(k_1,k_2,k'_3)$ 
give a negligible contribution to the integrations on {$k_1^-$} and on {$k_2^-$} and
 perform these integrations taking into account only the  poles 
of the quark propagators.
As a result the matrix elements of the current become the sum of : i) a triangle term (Fig. 1 (a)),
with the spectator quarks  on their
mass shell, and both the initial and the final nucleon vertexes in the valence sector,
and ii) a nonvalence term (Fig. 1 (b)), 
where the $q \bar q$ pair production appears and only the final 
nucleon vertex is in the valence sector. This latter term can be seen as an higher Fock state
contribution to the FF.
 Then, both the isoscalar and the isovector part of the quark-photon vertex ${\cal I}^\mu_3$
contain a purely valence contribution and a contribution corresponding to the pair
production (Z-diagram),
which can be decomposed in a
bare, point-like term and a vector meson dominance (VMD) term 
(according to  the decomposition of the photon
  state in bare,  hadronic  [and leptonic] contributions):
\begin{equation}
 {\cal I}^\mu_{i}(k,q) = {{\cal N}_{i}}~\theta(p^+-k^+)~\theta(k^+)
  \gamma^\mu+\theta({q}^+ + k^+)~
\theta(-k^+) \left [{Z^{i}_b ~{\cal N}_{i}} ~\gamma^\mu + Z^{i}_{V} \Gamma^\mu_{VMD}(k,q,i)\right]
\label{vert} 
\end{equation}
 with $i = IS, IV$ and ${\cal N}_{IS}=1/6$, ${\cal N}_{IV}=1/2$. 
The first term in (\ref{vert}) is the bare coupling of the triangle contribution, 
while $Z^{i}_b , ~Z^{i}_{V} $ are renormalization constants to be determined from
 the phenomenological analysis of the data.

\begin{figure}[htb]
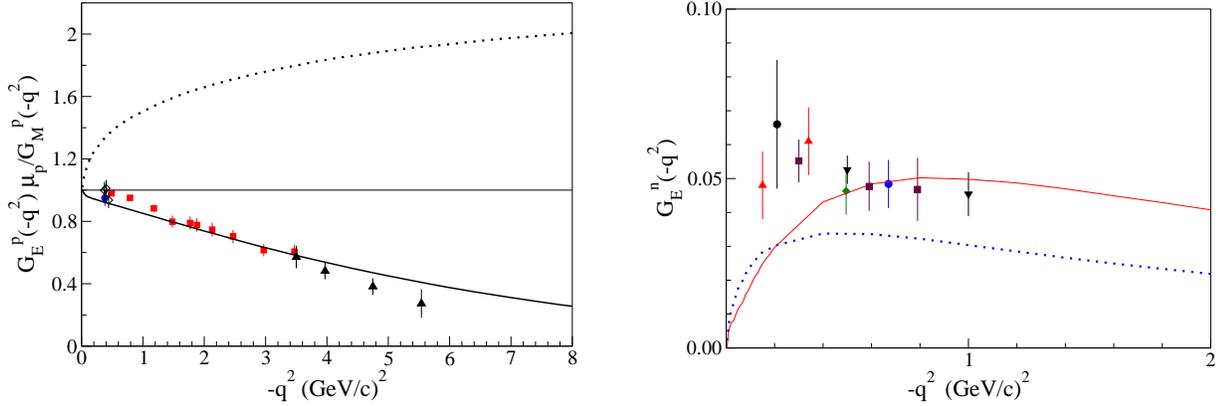

\begin{minipage}[t]{80mm}
\includegraphics[width=18pc]{FB18GepmupGmp.eps}
\label{fig:largenenough}
\end{minipage}
\hspace{\fill}
\begin{minipage}[t]{75mm}
\includegraphics[width=18pc]{FB18GEn.eps}
\label{fig:toosmall}
\end{minipage}
\caption{Nucleon electric form factors. Left panel: $\mu_p G^p_E(Q^2)/G^p_M(Q^2)$ vs $Q^2$. Right panel : $G^n_E(Q^2)$ vs
$Q^2$. Solid lines: full calculation, i.e., sum of triangle plus pair production terms. Dotted
lines: triangle contribution only. Data from \cite{JLAB2}.}
\end{figure}

 The  term $\Gamma^\mu_{VMD}(k,q,i)$ is obtained through the same microscopical VMD model
  already used in the pion case with the same VM eigenstates \cite{DFPS}.

{In the valence vertexes} the 3-momentum dependence  
is approximated through a
nucleon wave function a la Brodsky (PQCD inspired), namely
\be
\hspace{-0.6cm} {\cal W}_N\sim {(\xi_1\xi_2\xi_3)^{-0.12}
 \over \left[\beta^2 + M^2_0(1,2,3)\right]^3} 
\left[1 + A (M^2_0(1,2,3) - 9 m^2)~exp{\left (- (M^2_0(1,2,3) - 9 m^2)/2\beta^2 \right)} \right]
\ee
where $M_0(1,2,3)$ is the free mass of the three quark system. The parameter
$\beta $ is fixed through the nucleon magnetic moments, for which the values
$\mu_p =  2.878 ~ (Exp. ~2.793) $ 
and~ $\mu_n =  -1.859 ~(Exp. ~-1.913)$ are obtained.

In the non-valence vertex, needed to evaluate the Z-diagram
contribution, the momentum dependence  is approximated by
\be
\hspace{-0.9cm} G_N\sim {1 \over \left [\beta^2+M^2_0(1,2)\right]}~
\left \{ {1 \over \left [\beta^2+M^2_0(3',2)\right]} + 
{1 \over \left [\beta^2+M^2_0(3',1)\right]}\right \}  
\ee
with $M_0(i,j)$ the free mass of i and j quarks.

The  nucleon Sachs form factors can be otained from the matrix elements of the current by
means of proper traces. The preliminary results of Figs. 2 and 3 show clearly the 
relevance of the pair
production process and then of the nonvalence components for the nucleon FF.
 In particular the interplay of valence and nonvalence contributions generates
the possible zero in the proton electric form factor.

\begin{figure}[htb]
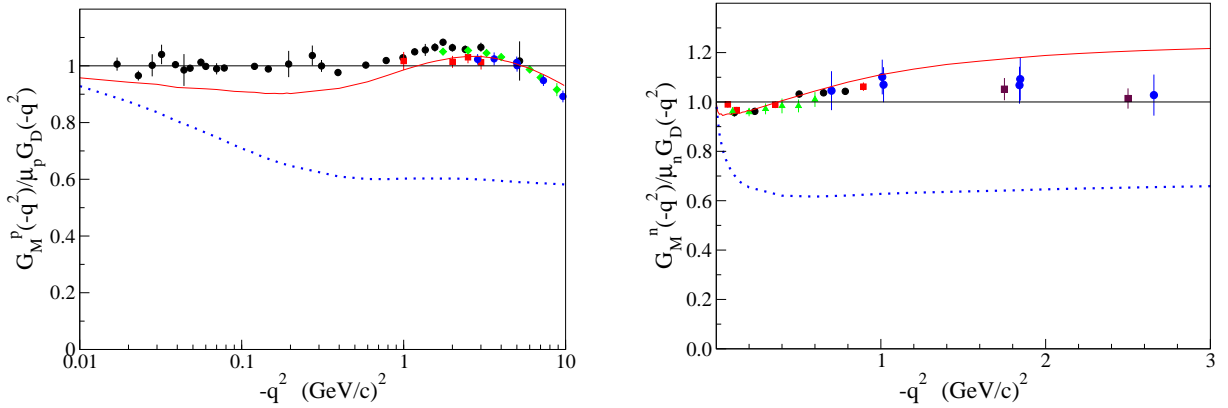

\begin{minipage}[t]{80mm}
\includegraphics[width=18pc]{FB18RGMpGD2.eps}
\label{fig:largenenough2}
\end{minipage}
\hspace{\fill}
\begin{minipage}[t]{75mm}
\includegraphics[width=18pc]{FB18RGMnGDext.eps}
\label{fig:toosmall2}
\end{minipage}
\caption{Nucleon magnetic form factors. Left panel: $G^p_M(Q^2)/(\mu_p G_D(Q^2))$ vs $Q^2$ with $G_D(Q^2)= 
[1 + Q^2/(0.71 (GeV/c)^2)]^{-2}$. 
Right panel : $G^n_M(Q^2)/(\mu_n G_D(Q^2))$ vs $Q^2$. Solid lines: full calculation, 
i.e., sum of triangle plus pair production terms. Dotted
lines: triangle contribution only. Data from \cite{JLAB2}.}
\end{figure}

\end{document}